\documentstyle[11pt,newpasp,twoside,epsf]{article}
\begin{document}
\title{Magnetic Fields in Molecular Clouds}
\author{Tyler L. Bourke$^1$ \& Alyssa A. Goodman$^2$}
\affil{
$^1$Harvard-Smithsonian Center for Astrophysics,
Submillimeter Array Project,
645 N.\ A'ohoku Place,
Hilo, HI 96720, USA\\
$^2$Harvard-Smithsonian Center for Astrophysics, 
60 Garden St, Cambridge MA 02138, USA}

\begin{abstract}

Magnetic fields are believed to play an important role in the evolution of
molecular clouds, from their large scale structure to dense cores,
protostellar envelopes, and protoplanetary disks.  How important is
unclear, and whether magnetic fields are the dominant force driving star
formation at any scale is also unclear.  In this review we examine the
observational data which address these questions, with particular emphasis
on high angular resolution observations.  Unfortunately the data do not
clarify the situation.  It is clear that the fields are important, but to
what degree we don't yet know.  Observations to date have been limited by
the sensitivity of available telescopes and instrumentation.  In the
future ALMA and the SKA in particular should provide great advances in
observational studies of magnetic fields, and we discuss which
observations are most desirable when they become available.

\end{abstract}

\section{Are Magnetic Fields Important?}

Magnetic fields are believed to play an important role in the evolution of
molecular clouds, and hence star formation.  However, despite progess on
both the observational and theoretical fronts in recent years, many
questions remain to be answered.  Fundamental questions include (1) what
is the dominant mechanism driving star formation, magnetic fields or
turbulence, and (2) how important are magnetic fields at different stages
in the star formation process?

For most of the past two decades the prevailing picture for the evolution
of a dense molecular cloud core to form a protostar has been one of
``quasi-static'' evolution of a strongly magnetised core through ambipolar
diffusion, over a time scale $>\!>$ the free-fall time, $t_{f\!f}$,
leading eventually to inside-out collapse onto the central region (Shu
et~al.\ 1987, 1999; Mouschovias \& Ciolek 1999). Recently a new theory has
emerged, where the molecular clouds themselves are short-lived phenomena
(liftimes a few $t_{f\!f}$ at most)  and the star formation process is
dynamical from the outset.  In this picture supersonic turbulence is the
dominant force in controlling the evolution of clouds and cores, and
regulates the star formation rate (see reviews by Mac Low \& Klessen 2003,
V\'azquez-Semadeni 2004, and reference therein).

The quasi-static model implies that cloud cores should be strongly
magnetically subcritical (i.e., static magnetic fields are strong enough
to provide support against gravity for $t >\!> t_{f\!f}$).  In addition to
supercritical cores (where the magnetic field provides no support), the
dynamical model is able to accommodate approximately critical or slightly
subcritical cores, as they will quickly evolve into supercritical cores
and collapse.  However, measurements of magnetic field strengths (and
hence magnetic flux-to-mass ratios) in cores, which are needed to
discrimate between the two models, are very difficult to obtain.  The only
practical tool available for directly measuring the field strength is the
Zeeman effect (Crutcher 1999, 2004).  Unfortunately molecules that are
sensitive to the Zeeman effect either have transitions that are too high
in frequency (e.g., CN at 113.5 GHz) or are too weak (CCS at 22-46 GHz)
for current instrumentation, or don't sample the densest gas in low-mass
cores (OH at 1.6 GHz), to provide a large set of measurements of the field
strength.

In the less dense regions of molecular clouds around the cores, where a
number of Zeeman measurements have been made (mainly using OH), both with
single dish radiotelescopes and interferometers, the results suggest that
clouds are approximately critical (Figure 1; Bourke et al. 2001; Crutcher
1999, 2004; see also Myers \& Goodman 1988), which does not seem to favour
the quasi-static model of magnetic field support.  However, measurements
of field strengths in dense cores ($n \geq 10^5$ cm$^{-3}$) are required.

\begin{figure*}[t]
\plotfiddle{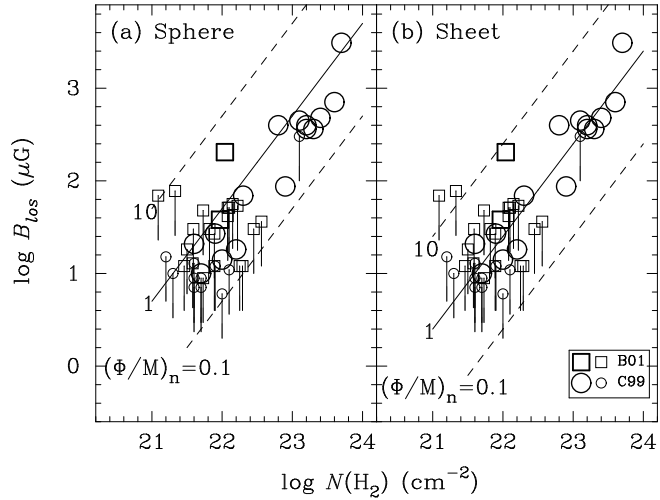}{6cm}{-90}{40}{40}{-150}{210}
\caption{Magentic field strengths and column densities from Bourke et al.\
2001 (B01) and Crutcher 1999 (C99).  See Bourke et al.\ 2001 for
details.  This plots indicates that Zeeman measurements 
are consistent with molecular clouds being in an approximately critical
state, ($\Phi/M)_n = 1$, or slightly supercritical ($<$ 1), depending on the 
adopted geometry.}
\end{figure*}

The direction and dispersion of polarization vectors tracing magnetic
field lines can theoretically be used to discriminate between the
quasi-static and turbulent models (Matthews et al.\ 2001, 2002, 2004;
Matthews \& Wilson 2002a,b;  Henning et al.\ 2001; Wolf et al.\ 2003;
Crutcher 2004).  If the field is sufficiently strong then a small
dispersion is predicted, and field lines should lie approximately parallel
to a dense core's minor axis, or show an hourglass morphology if it has
evolved toward the collapse phase, when observed with high angular
resolution.  In the turbulence driven model field lines may appear regular
on the larger size scales traced by the filaments in which the cores are
embedded, but are not expected to show any regular pattern on the size
scale of the cores themselves.  Of course, reality is never this simple.  
Observations of cores and filaments in polarized dust emission sometimes
show the regular pattern expected in the strong field case, but with
sufficient dispersion to imply that turbulence is important, and with mean
direction close to but not parallel to the cores' minor axes, as expected
in the quasi-static model (however, this last point can be explained as a
projection effect - Basu 2000).

At present the data do not strongly support either model.  In fact the
data can be explained with either model with a sufficient number of
caveats! Clearly real clouds and cores have both magnetic fields and
turbulence, and the dominant mechanism may be region dependent, e.g., MHD
turbulence on large scales and ambipolar diffusion on small scales.  
Until a large number of magnetic field measurements are made in the
densest regions of cores, and the full three dimensional field structure
of clouds and cores can be determined, deciding between the two will be
difficult.

In the following sections we look in more detail at the observations that
lead us to these conclusions, in particular those made with ``high angular
resolution'', which for this field does not automatically imply
interferometers.  We also examine what future observations are required to
make progress in this field.

\section{Observational Overview - High Resolution Observations}

Due to the difficultly in making magnetic field measurements with current
interferometers, this section on high resolution observations of magnetic
fields includes Zeeman and dust polarization measurements made with
sub-arcminute beams on single dish telescopes (i.e., IRAM 30-m; JCMT
15-m).

Magnetic fields in molecular clouds can be probed by a number of means.  
The Zeeman effect has been used to determine the line-of-sight field
strength ($B_{los}$)  in thermal lines, both in emission and absorption
(Crutcher 1999; Bourke et al.\ 2001; and references therein). Masers are
also potential tools, but the uncertainty in determining the physical
conditions and sizes of the masing regions make it difficult to interpret
the results.  Polarimetry of aligned dust grains is the other major
observational technique, which can be used to map the morphology of the
field in the plane of the sky, and to estimate the field strength ($B_p$)
using the modified Chandrasekhar-Fermi (C-F) method (Ostriker et al.\
2001; Padoan et al.\ 2001; Heitsch et al.\ 2001a).  Studies have been
conducted in both the near-infrared, the far-infrared, and more recently
in the (sub)millimetre. As discussed in the following sections, each
method has a number of shortcomings and full information on the 3D
structure of magnetic fields in molecular clouds are lacking in most
cases.  Possible methods for obtain such information are discussed in \S3.

\subsection{Zeeman Measurements}

Most observations of the Zeeman effect in molecular clouds have been
performed using single dish telescopes observing the OH transitions at 1.6
GHz, with a few studies using the VLA.  Other secure detections of the
Zeeman effect have come from HI, CN and excited transitions of OH
(G\"usten et al.\ 1994).

High resolution Zeeman observations of the densest regions of nearby low
mass cores are almost non-existent.  CCS observations of the chemically
evolved, heavily depleted starless core L1498 (Levin et al.\ 2001) and the
chemically young L1521E (Shinnaga et al.\ 1999; see also Aikawa 2004) have
been made with sub-arcminute resolution. Shinnaga et al.\ claim a
detection with $B_{los} = 160\pm46 \mu$G using the 45 GHz transition, and
state that this value is larger than the critical value (which
unfortunately they do not give).  This result requires confirmation,
preferably using other CCS transitions at lower frequencies.

The Zeeman effect has been observed in the CN 1-0 transition at 113 GHz
with the IRAM 30-m (23$''$ beam) by Crutcher et al.\ (1999) toward four
cores associated with the high mass star-forming regions OMC1, DR21OH and
M17SW.  As the {\em frequency offset} due to the Zeeman effect is {\em
independent} of the line frequency, whereas the {\em Doppler broadened
line width} is {\em proportional} to the line frequency, the ratio of the
Zeeman effect to the line width decreases as the frequency of the line
increases.  This makes observations of the Zeeman effect at mm wavelengths
much less sensitive than at cm wavelengths, hence the paucity of
observations with this potentially rewarding transition.  Because CN 1-0
consists of a number of hyperfine components with different Zeeman
effects, the detections are fairly secure.  Crutcher et al.\ conclude that
the cores observed are supercritical by a factor 2--3.

VLA Zeeman observations with beam sizes $5-60''$ in the HI 21 cm and OH 18
cm transitions have been performed toward a number of high-mass star
forming regions (e.g., W3 -- Roberts et al.\ 1993; S106 -- Roberts et al.\
1995;  DR21 -- Roberts et al.\ 1997; NGC 2024 -- Crutcher et al.\ 1999;
M17 -- Brogan et al.\ 1999, 2001; NGC 6334 -- Sarma et al.\ 2000).  In
many cases it is claimed that the HI and/or OH emission is in fact tracing
high density ($>10^4$ cm$^{-3}$) gas in a PDR interface between the
ionized and molecular regions.  Line-of-sight field strengths up to 0.5 mG
have been observed in most regions, with large variations across the
mapped areas (Figure 2). In NGC 6334 the field actually changes sign,
while in other regions it drops to 0 in places, indicating significant
changes in direction of the field.  This result, and the similar result in
M17, could explain the lack of detection in single dish Zeeman
observations in these sources (Bourke et al.\ 2001).  In all these studies
it is inferred that the magnetic field is either approximately critical
(W3, S106, NGC 6334) or supercritical by a factor 2--3 (NGC 2024, M17).  
There is certainly no clear evidence for a subcritical cloud in any
observations of the Zeeman effect alone.

\begin{figure*}[t]
\plotfiddle{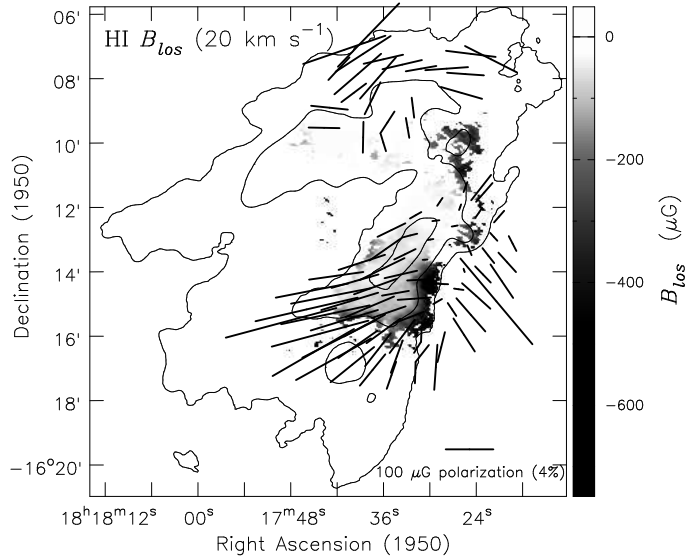}{6.3cm}{0}{50}{50}{-150}{-90}
\caption{Greyscale image of the line-of-sight field strength $B_{los}$
measured in the 20 km/s HI component toward M17SW with the VLA (Brogan et
al.\ 2001).  The contours represent the 21 cm continuum emission.
The magnetic field direction in the plane of the sky is indicated 
with the superimposed 100 $\mu$m polarization vectors, rotated 90\deg\
(Dotson et al.\ 2000).}
\end{figure*}

\subsection{Dust and Spectral Line Polarization}

Some years ago it was hoped that polarized background starlight observed
in the infrared would be a useful probe of the denser regions of molecular
clouds, but the percentage polarization was not observed to increase as
expected (Goodman et al.\ 1995).  More recently the polarized thermal
emission in the far-infrared and (sub)millimetre regions has been used to
infer the field direction in the plane of the sky from a number of clouds
and cores (Dotson et al.\ 2000; Davis et al.\ 2000;  Matthews et al.\
2001, 2002, 2004; Matthews \& Wilson 2002a,b;  Henning et al.\ 2001; Wolf
et al.\ 2003; Crutcher et al.\ 2004). Interestingly these studies also
find that the percentage polarization does not increase toward the regions
of maximum intensity (and hence density).  In fact a {\em decrease} in
percentage polarization is observed in most cases.  The likely explanation
in most cases is poor grain alignment due to spherical grain growth (``bad
grains'' -- Goodman et al.\ 1995; Lazarian et al.\ 1997; Padoan et al.\
2001).

The far infrared observations of high mass star forming regions using the
KAO have been discussed elsewhere (Dotson et al.\ 2000 and references
therein).  All the KAO maps show regular structures (but not necessarily
uniform) with depolarization at the highest intensities.  It would be
useful to apply more recent analysis techniques to these data, for example
the modified C-F method.

Two instruments have been used to obtain most of the (sub)millimetre
results published at this time -- the SCUBA polarimeter at 850 $\mu$m, and
BIMA at 3 and 1 mm.  Observations have also been made with OVRO (Akeson \&
Carlstrom 1997) and the 350 $\mu$m polarimeter HERTZ on the CSO
(Schleuning et al.\ 2000; Houd\'e et al.\ 2002).

Matthews et al.\ (2004) have combined SCUBA and BIMA observations of Orion
and Perseus (B1), comparing the large and small scale field directions.
They find that in Orion the field direction in the cores is similar to
that of the filaments in which they are embedded, but in B1 the cores show
different orientations.  The reason for these differences is unclear, but
they could imply that B1 is globally supported by turbulence, with local
density enhancements able to undergo collapse, whereas Orion is not
turbulently supported (although its turbulence is ``greater''), resulting
in more ordered field lines on all scales (Heitsch et al.\ 2001b;  Mac Low
\& Klessen 2003).  The relevant physical parameters need to be evaluated
to examine this (e.g., mass-to-flux ratio, turbulent line width, virial
terms).  A SCUBA map of the Serpens region, containing a number of
protostars, was presented by Davis et al.\ (2000).  In the NW cluster the
field shows some degree of regularity, but in the presumably older SE
there is a large dispersion in field direction between the protostars,
possibly suggesting the field becomes less important at the core size
scale as star formation progresses.

Recent mapping studies of individual low mass starless cores
(Ward-Thompson et al.\ 2000; Crutcher et al.\ 2004) and protostars
(Henning et al.\ 2001; Wolf et al.\ 2003; Valle\'e et al.\ 2003) have been
made with SCUBA.  All these results show depolarization at the highest
intensities (Padoan et al.\ 2001).  In starless cores the fields are
uniform, but not aligned with the core minor axes, displaying offsets of
up to 30\deg, and all cores are found to be supercritical (or at least are
not clearly subcritical), with field strengths inferred using the modified
C-F method ($B_p \sim$50-150 $\mu$G).  In protostellar cores the fields do
not appear to be as uniform (possibly due to outflow disruption), and no
clear preferred orientation with respect to either the outflows or cores
is evident, although there is some suggestion the field lines are either
aligned parallel or perpendicular to the outflow axis on a case by case
basis. Field strengths estimated via the modified C-F method are typically
100--200 $\mu$G, but unfortunately are not compared to the critical
values.

Interferometric studies at mm wavelengths have been made with both OVRO
(Akeson \& Carlstrom 1997) and BIMA (Rao et al.\ 1998; Lai et al.\ 2002,
2003a), of both low and high mass protostars. As the OVRO observations
only produced a couple of measurements per field, we concentrate on the
BIMA results.  Rao et al.\ (1998) observed Orion-KL at 3 and 1 mm with
$\sim$5$''$ beams.  A relatively ordered field was observed except near
the position of IRc2, where the field direction changed by 90\deg.  This
is explained as the effect of the outflow on the dust grains causing
alignment due to streaming motions (the ``Gold'' effect -- Lazarian 1997).
Lai et al.\ (2002) observed NGC 2024 FIR 5 at 1 mm with 2$''$ resolution.
A uniform field was observed, with a slow change in position angle, which
they modelled as due to an hourglass shaped morphology.  They claim the
pattern is due to contraction of the psuedo-disk perpendicular to the
field (e.g., Galli \& Shu 1993).  Applying the modified C-F method they
infer a field strength in the plane of the sky of $\sim$3.5 mG.  This is
extremely large compared to the Zeeman OH result of 65 $\mu$G for the
line-of-sight component.  If the field is not close to the plane of the
sky, as these numbers would suggest, then the result could be explained as
due to beam averaging of the OH data (60$''$) or the OH data does not
trace the same high density region as the 1 mm dust emission.  Another
explanation is the modified C-F method is not applicable in this case.

\begin{figure}[t]
\plotfiddle{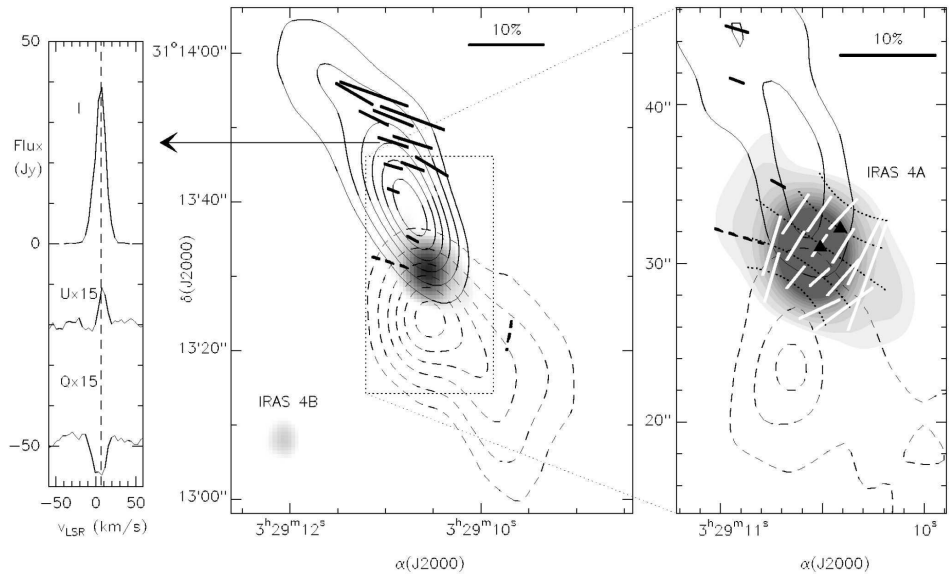}{7.0cm}{0}{80}{80}{-180}{0}
\caption{{\em Middle panel:} The greyscale shows the 1.3 mm dust emission
from NGC 1333 IRAS 4 as observed with BIMA by Girart et al.\ (1999).
Superimposed are contours of redshifted (solid) and blueshifted (dashed) 
integrated intensity of CO 2-1, and vectors showing the position angle of
linearly polarized CO 2-1 emission.  {\em Right panel:} A magnified map of the
central region, with white vectors indicating the position angle of the
dust polarization, and the dotted lines indicating a possible pinch
magnetic configuration.  {\em Left panel:} CO 2-1 spectra of Stokes I, U and Q
averaged over the redshifted outflow lobe, with the cloud ambient velocity
indicated.}
\end{figure}

Linear polarization of CO emission has been observed with BIMA toward the
low mass protostar NGC 1333 IRAS 4A (Girart et al.\ 1999; Figure 3), and
the high mass protostellar region DR21(OH) (Lai et al.\ 2003a).  Girart et
al.\ find that the dust polarization pattern toward IRAS 4A is consistent
with a pinch (hourglass) configuration.  Linear polarization of CO is
detected mainly toward the redshifted outflow lobe.  The polarization of
spectral lines is predicted to be either parallel or perpendicular to the
field, depending on a number of factors (Goldreich \& Kylafis 1982).  
Girart et al.\ argue that in this case the field is parallel to the
polarization vectors (which are perpendicular to the dust polarization
vectors), and speculate that the magnetic field is bending the outflow
(Figure 3).  In the high mass protostar DR21(OH) Lai et al.\ also infer
that the field traced by the linearly polarized CO emission is parallel to
the field, by comparison with the polarized dust emission.  Polarized CO
emission is observed over a larger area than the dust, allowing the field
morphology to be inferred over a similar area.  Lai et al.\ deduce that
the two dust continuum peaks (MM1 \& MM2) are condensations within a
magnetic flux tube. Applying the modified C-F method field strengths of
$B_p \sim 1$ mG are inferred, about twice that found for $B_{los}$ through
CN Zeeman observations (Crutcher et al.\ 1999).  Combining these results
implies that the field is pointed toward us at an angle of $\sim$30\deg to
the line of sight.

\section{Probing the Magnetic Field in 3D}

For many years observational studies of magnetic fields in molecular
clouds were restricted to probing the line-of-sight component via the
Zeeman effect in thermal (non-maser) lines, or the plane of the sky
component via dust polarization (and more recently linearly polarized
spectral lines).  A method to combine Zeeman observations with polarized
background starlight was proposed by Myers \& Goodman (1991; see also
Goodman \& Heiles 1994) to deduce the 3D field structure. However, this
method has not been commonly used, perhaps a result of concerns about the
usefulness of background polarized starlight in molecular clouds
(depolarization; lack of stars), and the lack of Zeeman observations over
the large angular sizes required at that time to obtain a sufficiently
large number of stars for polarization studies. Recent theoretical
simulations have shown that a modified C-F method can be used to infer the
plane of the sky field strength under certain conditions (see Ostriker et
al.\ 2001; Padoan et al.\ 2001; Heitsch et al.\ 2001a for specifics), and
when applying this method to observational data care must be taken to
ensure these conditions are met. If the field probed by the Zeeman effect
and the dust polarization is believed to arise from a common region, then
combining the Zeeman and C-F methods enables the full field strength and
its angle to be determined. This has been attempted by Lai et al.\ (2003b)
for DR21(OH) using CN 1-0 Zeeman data (Crutcher et al.\ 1999).  
Unfortunately the densities probed by the dust continuum ($>10^5$
cm$^{-3}$) are not in general probed by OH, the molecular most used in
Zeeman observations.  Potential Zeeman sensitive molecules that probe
these densities (CN, CCS, CCH, SO) are discussed later.

Another technique for probing the 3-D field structure in the weakly
ionized regions of molecular clouds (i.e., dense cores) involves the use
of Zeeman data, dust polarimetry, and measurements of the ratio of
ion-to-neutral line widths (Houd\'e et al.\ 2002).  The advantage of this
technique over the simple one describe above is that all the quantities
needed are measured directly by observations, unlike the modified C-F
method. Houd\'e et al.\ used this method to infer the structure of the
field in M17, and more recently Lai et al.\ (2003b) have used the
technique to infer the field in DR21(OH) at high angular resolution,
combining their BIMA data described above with new OVRO data.  Although
Lai et al. find that the angle of the field to our line of sight is
unchanged from their earlier estimate, the field strength is significantly
lower (0.4 mG cf. 1 mG), which they suggest is a result of overestimating
the field strength using the modified C-F method, due to smoothing of the
polarization dispersion in the BIMA data.

With the next generation of interferometers combining great improvements
in sensitivity with high angular resolution (ALMA, SKA), it may be
possible to use these techniques to determine the full field structure and
strengths in a more representative sample of molecular clouds and cores.

\section{Unanswered Questions \& Future Directions}

As stated at the beginning of this review there are two fundamental
questions into which we would like to gain insight: (1) what is the
dominant mechanism driving star formation, and (2) how important are
magnetic fields at different stages in the star formation process?

The results discussed here unfortunately are inconclusive to answer (1).
Most of the Zeeman observations and many of the polarization studies have
been toward regions that are already forming stars, where the magnetic
field should not be dominant.  So it is no surprise that this is the
result found through observations.  In the few observations of starless
cores, the observations again suggest the field is important though not
dominant (Crutcher et al.\ 2004).  However, the isolated starless cores
L183 and L1544 show spectral signatures of inward motions (Lee et al.\
2001), and so appear to be at an advanced stage of evolution just prior to
collapse.

For core evolution most of the results suggest that by the time the
protostellar stage is reached magnetic fields are not energetically
dominant.  But they are still important, as shown by the many
observational examples of uniform and ordered polarization patterns, and
in some cases the hourglass-like morphologies which might suggest core
contraction due to ambipolar diffusion, as least during the inital stages
of evolution toward forming a protostar.

At present the observations are insufficient to address (2).  The
observations of high mass regions suffer either from lack of resolution,
even with interferometers, or don't sample the diffuse parts of molecular
clouds which are more representative of the overall cloud than those
regions that have obtained sufficient density to form stars and are
therefore bright enough to be well detected by Zeeman or polarization
studies.  Studies of low mass regions suffer from similar problems, and in
addition do not probe every evolutionary stage, from chemically young
protostellar cores (Aikawa 2004), through to protoplanetary disks (Dutrey
2004; Wilner 2004).

New instruments will help us to obtain a little more knowledge on both
these issues.  In particular we highlight the importance of ALMA in dust
polarization and linearly polarized emission line studies at (sub)mm
wavelengths, and the SKA for Zeeman studies.

In order to make progress on question (1) we need Zeeman measurements
throughout molecular clouds, which could be provided by OH Zeeman
observations with the SKA of lines which are too weak to provide
detections with existing telescopes.  If the modified C-F method can be
tested more thoroughly in simulations and if it is applied correctly to
observational data, it may be a useful tool to provide information on the
3D field when used with Zeeman data.  These observations will not be easy.
The technique of Myers \& Goodman (1991) to combine Zeeman and background
polarized starlight observations to infer the 3D field structure should be
re-examined as a tool to probe the lower density regions of molecular
clouds (which contain the bulk of the material), particularly with today's
10-m class optical telescopes and infrared array cameras.

We would like to know the field strengths within dense protostellar cores
before the onset of star formation.  Observations using ALMA of CN 1-0 at
113.5 GHz, CCH at 85 GHz, and SO at 30 and 100 GHz, and the SKA of CCS at
11 and 22 GHz (Bel \& Leroy 1989, 1998; Shinnaga \& Yamamoto 2000), may
provide these data, particular if the cores are carefully selected so that
these molecules are not selectively depleted (e.g., L1521E).  In such
cores the method of Houd\'e et al.\ may provide information on the full 3D
structure of the field.  We would also like to know this information for
protostellar cores at both the Class 0 and Class I phases, to examine the
importance of the field during protostellar evolution.

In order to understand the high resolution observations of dust
polarization, the observed depolarized at high intensities needs to be
explained quantitatively.  Further, numerical simulations of turbulence
dominant molecular clouds that include magnetic fields need to be able to
resolve protostellar cores and their fields, and not just treat them as
sink particles, for comparison with observations (V\'azquez-Semadeni
2004).

The new generation of interferometers (CARMA, ALMA and SKA) will provide
the sensitivity and resolution required to make progress on two questions
where essentially no observational information exists on the magnetic
field -- how important are magnetic fields in protostellar disks (disk
viscosity, angular momentum transport; Hartmann 1998), and what drives and
collimates protostellar outflows and jets (X-wind -- Shang 2004; Disk wind
-- K\"onigl \& Pudritz 2000).  Zeeman observations of some or all of CN,
CCH, CCS, and SO in thermal emission will be very important in the study
of protostellar disks (Qi 2000; Dutrey 2004) and envelopes (van Dishoeck
\& Blake 1998;  Aikawa 2004), in particular if the full field can be
inferred by combining these data with linear polarization studies of dust
and CO. Polarization observations at size scales of a few AU or better may
help to decide which mechanism is responsible for launching and
collimating protosteller jets.  These observations will still be difficult
even with ALMA et al.

So our final words are directly to those planning the construction of ALMA
and the SKA:
\begin{itemize}
\item Please provide ALMA with good polarimeters for dust and spectral line
polarization studies at submm wavelengths, and the capability 
for Zeeman observations down to 30 GHz.
\item Please push the upper frequency of the SKA to at least 24 GHz, to
allow for Zeeman observations of thermal CCS (22 GHz), of water masers (22
GHz), and non-Zeeman observations of the important inversion transitions of
ammonia near 24 GHz.  Please pay particular attention to the polarization
characteristics of potential array designs.  
\end{itemize}
THANKS!


\begin{references}

\reference Aikawa, Y., 2004, in IAU Symp. 221, Star Formation
at High Angular Resolution, ed.\ M.G.\ Burton, R.\ Jayawardhana 
\& T.L.\ Bourke (PASP: San Francisco), in press

\reference Akeson, R.L., \& Carlstrom, J.E., 1997, ApJ, 491, 254

\reference Basu, S., 2000, ApJ, 540, L103

\reference Bel, N., \& Leroy, B., 1989, A\&A 224, 206

\reference Bel, N., \& Leroy, B., 1998, A\&A 335, 1025

\reference Bourke, T.L., Myers, P.C., Robinson, G., \& Hyland, A.R., 2001,
ApJ, 554, 916

\reference Brogan, C.L., Troland, T.H., Roberts, D.A., \& Crutcher, R.M.,
1999, ApJ, 515, 304

\reference Brogan, C.L., Troland, T.H., 2001, ApJ, 560, 821

\reference Crutcher, R.M., 1999, ApJ, 520, 706

\reference Crutcher, R.M., 2004, in The Magnetized Interstellar Medium, ed.
B.\ Uyansker, W.\ Reich, \& R.\ Wielebinski, in press

\reference Crutcher, R.M., Troland, T.H., Lazareff, B., Paubert, G., \&
Kaz\`es, I., 1999a, ApJ, 514, L121

\reference Crutcher, R.M., Roberts, D.A., Troland, T.H., \& Goss, W.M.,
1999b, ApJ, 515, 275

\reference Crutcher, R.M., Nutter, D.J., Ward-Thompson, D., \& Kirk, J.M.,
2004, ApJ, in press

\reference Davis, C.J., Chrysostomou, A., Matthews, H.E., Jenness, T., \&
Ray, T.P., 2000, ApJ, 530, L115

\reference Dotson, J.L., Davidson, J., Dowell, C.D., Schleuning, D.A., \&
Hilderbrand, R.H., 2000, ApJSS, 128, 335

\reference Dutrey, A., 2004, in IAU Symp. 221, Star Formation
at High Angular Resolution, ed.\ M.G.\ Burton, R.\ Jayawardhana
\& T.L.\ Bourke (PASP: San Francisco), in press

\reference Galli, D., \& Shu, F.H., 1993, ApJ, 417, 220

\reference Girart, J.M., Crutcher, R.M., \& Rao, R., 1999, ApJ, 525, L109

\reference Goldreich, P., \& Kylafis, N.D., 1982, ApJ, 253, 606

\reference Goodman, A.A., \& Heiles, C., 1994, ApJ, 424, 208

\reference Goodman, A.A., Jones, T.J., Lada, E.A., \& Myers, P.C., 1995,
ApJ, 448, 748

\reference G\"usten, R., Fiebig, D., \& Uchida, K.I., 1994, A\&A, 286, L51 

\reference Hartmann, L., 1998, Accretion Processes in Star Formation
(Cambridge Univ. Press).

\reference Heitsch, F., Zweibel, E.G., Mac Low, M.-M., Li, P.S., Norman,
M.L., 2001a, ApJ, 561, 800

\reference Heitsch, F., Mac Low, M.-M., \& Klessen, R.S., 2001b, ApJ, 547,
280

\reference Henning, Th., Wolf, S., Launhardt, R., \& Waters, R., 2001, ApJ,
561, 871

\reference Houde, M.~et al.\ 2002, ApJ, 569, 803 

\reference K\"onigl, A., \& Pudritz, R.F., 2000, in Protostars \& Planets
IV, ed. V.\ Mannings, A.P.\ Boss, S.S.\ Russell (Tucson: Univ. Arizona
Press), p.\ 759

\reference Lai, S.-P., Crutcher, R.M., Girart, J.M., \& Rao, R., 2002, ApJ,
566, 925

\reference Lai, S.-P., Girart, J.M., Crutcher, R.M., 2003a, ApJ, 598, 392

\reference Lai, S.-P., Velusamy, T., \& Langer, W.D., 2003b, ApJ, 596, L239

\reference Lazarian, A., 1997, ApJ, 483, 296

\reference Lazarian, A., Goodman, A.A., \& Myers, P.C., 1997, ApJ, 490, 273

\reference Lee, C.W., Myers, P.C., \& Tafalla, M., 2001, ApJS, 136, 703 

\reference Levin, S.M., Langer, W.D., Velusamy, T., Kuiper, T.B.H., 
\& Crutcher, R.M., 2001, ApJ, 555, 850 

\reference Mac Low, M.-M., \& Klessen, R.S., 2003, Rev. Modern Physics, in
press (astro-ph/0301093)

\reference Matthews, B.C., Wilson, C.D., \& Fiege, J., 2001, ApJ, 562, 400

\reference Matthews, B.C., Fiege, J., \& Moriarty-Schieven, G., 2002, ApJ,
569, 304

\reference Matthews, B.C., \& Wilson, C.D., 2002a, ApJ, 571, 356

\reference Matthews, B.C., \& Wilson, C.D., 2002b, ApJ, 574, 822

\reference Matthews, B.C., Lai, S.-P, Crutcher, R.M., Wilson, C.D., 2004, 
in IAU Symp. 221, Star Formation
at High Angular Resolution, ed.\ M.G.\ Burton, R.\ Jayawardhana 
\& T.L.\ Bourke (PASP: San Francisco), in press

\reference Mouschovias, T.Ch., \& Ciolek, G.E., 1999, in The Origin of
Stars and Planetary Systems, ed. C.J.\ Lada \& N.D.\ Kylafis (Kluwer:
Dordrecht), p.\ 305

\reference Myers, P.C., \& Goodman, A.A., 1988, ApJ, 326, L27

\reference Myers, P.C., \& Goodman, A.A., 1991, ApJ, 373, 509

\reference Ostriker, E.C., Stone, J.M., \& Gammie, C.F., 2001, ApJ, 546, 980

\reference Padoan, P., Goodman, A.A., Draine, B.T., Juvela, M., Nordlund,
\AA., \& R\"ognvaldsson, \"O.E., 2001, ApJ, 559, 1005

\reference Qi, C., 2000, PhD Thesis, Caltech.

\reference Rao, R., Crutcher, R.M., Plambeck, R.L., \& Wright, C.H., 1998,
ApJ, 502, L75

\reference Roberts, D.A., Crutcher, R.M., Troland, T.H., \& Goss, W.M.,
1993, ApJ, 412, 675

\reference Roberts, D.A., Crutcher, R.M., Troland, T.H., 1995, ApJ, 442, 208

\reference Roberts, D.A., Dickel, H.R., \& Goss, W.M., 1997, ApJ, 476, 209

\reference Sarma, A.P., Troland, T.H., Roberts, D.A., \& Crutcher, R.M.,
2000, ApJ, 533, 271

\reference Schleuning, D.A., Vaillancourt, J.E., Hilderbrand, R.H., Dowell,
C.D., Novak, G., Dotson, J.L., \& Davidson, J.A., 2000, ApJ, 535, 913

\reference Shang, H., 2004, in IAU Symp. 221, Star Formation
at High Angular Resolution, ed.\ M.G.\ Burton, R.\ Jayawardhana
\& T.L.\ Bourke (PASP: San Francisco), in press 

\reference Shinnaga, H., Tsuboi, M., \& Kasuga, T., 1999, in 
Proceedings of Star Formation 1999, ed.\ T.Nakamoto
(Nobeyama Radio Observatory), p.\ 675 

\reference Shinnaga, H., \& Yamamoto, S., 2000, ApJ, 544, 330

\reference Shu, F.H., Adams, F.C., \& Lizano, S., 1987, ARAA, 25, 23

\reference Shu, F.H., Allen, A., Shang, H., Ostriker, E.C., \& Li, Z.-Y.,
1999, in The Origin of Stars and Planetary Systems, ed.\ C.J.\ Lada \& 
N.D.\ Kylafis (Kluwer: Dordrecht), p.\ 193

\reference Valle\'e, J.P., Greaves, J.S., \& Fiege, J.D., 2003, ApJ, 588, 910

\reference van Dishoeck, E.F., \& Blake, G.A., 1998, ARAA, 36, 317

\reference V\'azquez-Semadeni, E., 2004, in IAU Symp. 221, Star Formation
at High Angular Resolution, ed.\ M.G.\ Burton, R.\ Jayawardhana 
\& T.L.\ Bourke (PASP: San Francisco), in press

\reference Ward-Thompson, D., Kirk, J.M., Crutcher, R.M., Greaves, J.S.,
Holland, W.S., \& Andr\'e, P., 2000, ApJ, 537, L135

\reference Wilner, D.J., 2004, in IAU Symp. 221, Star Formation
at High Angular Resolution, ed.\ M.G.\ Burton, R.\ Jayawardhana
\& T.L.\ Bourke (PASP: San Francisco), in press

\reference Wolf, S., Launhardt, R., \& Henning, T., 2003, ApJ, 592, 233

\end{references}
\end{document}